\documentclass[12pt]{article}
\usepackage{graphicx}
\usepackage[bindingoffset=0cm,left=2cm,top=2.5cm,bottom=2.5cm,right=2.5cm]{geometry}

\begin{document}

\begin{center}
{\large \bf QUBIT PORTRAIT \\ OF THE PHOTON-NUMBER TOMOGRAM \\ AND
SEPARABILITY OF TWO-MODE LIGHT STATES\\[10mm]}
Sergey N. Filippov$^{1}$ and Vladimir I. Man'ko$^{2}$
\\[8mm]
\textit{$^{1}$ Moscow Institute of Physics and Technology (State University)\\
Institutskii per. 9, Dolgoprudnyi, Moscow Region 141700,
Russia
\\[1mm]
$^{2}$ P. N. Lebedev Physical Institute, Russian Academy of Sciences\\
Leninskii Prospect 53, Moscow 119991, Russia
\\[1mm]}
e-mail: filippovsn@gmail.com \ manko@sci.lebedev.ru
\\[3mm]
\end{center}

\begin{abstract}
In view of the photon-number tomograms of two-mode light states,
using the qubit-portrait method for studying the probability
distributions with infinite outputs, the separability and
entanglement detection of the states are studied. Examples of
entangled Gaussian state and Schr\"{o}dinger cat state are
discussed.
\end{abstract}


\section{\label{introduction}Introduction}

The probability representation of quantum states with continuous
variables was suggested in \cite{tombesi-manko}. According to this
representation, any quantum states is associated with a fair
probability-distribution function called symplectic tomogram (or
tomographic-probability distribution). Information contained in
the tomographic-probability distribution is the same that is in
the density operator. The optical tomograms \cite{berber,vogel}
were used earlier as a technical tool to reconstruct the Wigner
function \cite{wigner} which was identified with the quantum
state. The experiments to measure quantum states \cite{raymer} are
aimed at obtaining the Wigner function by means of measuring the
homodyne quadrature component. Though this procedure implies the
measurement of optical tomograms, the latter ones were not
identified with the quantum states in these experiments. The
optical tomogram is considered as intermediate technical
information that gives the opportunity to "reconstruct" the state
identified with the Wigner function.

\bigskip

An understanding of the fact that tomograms (optical or symplectic
ones) are themselves a primary notion of quantum states was
suggested in \cite{tombesi-manko} (see also
\cite{mendes-physica,oman'ko-97,marmoJPA,sudarshan-2008}). The
analogues treatment of quantum states of spin systems (qubits,
qudits) as tomographic-probability distributions was proposed in
\cite{dodonovPLA,oman'ko-jetp}. Thus, in the probability
representation of quantum mechanics, which is completely
equivalent to other representations like Schr\"{o}dinger
representation \cite{schrodinger}, Feynmann path integral
representation \cite{feynmann}, Moyal representation \cite{moyal},
etc., the quantum states are described by the standard probability
distributions (state tomograms). Spin tomograms were used to
discuss the separability and entanglement phenomena in
\cite{lupo1,lupo2,lupo3}. The qubit-portrait method for studying
entanglement of multiqudit systems was suggested in
\cite{chernega}. This method is similar to the integration method
of symplectic tomograms applied in \cite{shchukin} (see also
\cite{kolesnikov-mipt}), with both methods providing the analogs
of two-qubit states.

\bigskip

There exist descriptions of the photon states in terms of the
photon-number tomograms \cite{banaszek,wvogel,mancini}. The
photon-number tomogram is the probability distribution of a
discrete variable $n=0,1,2,...$. This distribution function also
depends on an extra complex parameter and contains complete
information on the quantum state. The photon-number tomograms for
the Gaussian states were constructed in \cite{omanko}. The aim of
our work is to develop an analog of the qubit portrait method that
was suggested for qudit states and apply this method to the
photon-number tomogram of multimode photon systems. This means
that we consider a linear map of discrete probability
distributions with infinite probability vectors onto probability
distributions with finite probability vectors. The map, which is
applied to the photon-number tomograms, is not a positive map
(see, e.g., \cite{sudarshan}). Nevertheless, it provides a
possibility to detect the entanglement of multimode photon states.
Positive maps were used to detect entanglement in
\cite{horodecki,peres,simon} and nonpositive maps were used to
detect entanglement, e.g., in
\cite{sudarshan-witherr,sudarshan-withouterr,chirkin-1,chirkin-2,chirkin-3}.

\bigskip

In this paper we focus on two-mode nonclassical states of light,
which are of great interest since they can have some particular
features \cite{dodonov-review,book}. Entanglement
\cite{schrodinger} of quantum states turned out to be a very
important tool for secure quantum communication and quantum
cryptography. A geometrical interpretation of entangled two-qudit
states was proposed in \cite{filipp}. In spite of the fact that
detection of entanglement has undergone rapid development in the
last few decades, they do not cover all the possibilities. The
main goal of the present article is to introduce an alternative
method that can reveal the entanglement of two-mode light states.
There have been already made successful attempts to solve this
problem \cite{simon,illuminati}, for instance, by using the
Bell-CHSH inequality \cite{bell,chsh} within the framework of
symplectic tomograms \cite{tombesi-manko,shchukin,d'ariano}. Here,
we try to attack the problem with the help of the photon-number
tomogram \cite{banaszek,wvogel,mancini} which provides additional
information on quantum correlations.

\bigskip

Since the entanglement of two qubit system has been sufficiently
investigated, it looks feasible to detect the entanglement of the
system involved by using a linear map of the photon-number
tomogram with infinite outputs onto qubit tomogram (qubit
portrait). The qubit-portrait method to study qudit states was
introduced in \cite{chernega} and developed in \cite{lupo1}. As
far as there exist many ways to construct a qubit portrait, we are
going to discuss some examples of them. The procedure above
results in reducing the separability property of a two-mode light
state to the Bell-CHSH inequality for two qubits. That inequality
is fulfilled if the initial two-mode light state is separable. It
naturally leads to a necessary condition of separability. Its
violation indicates immediately that the state in question is
entangled.

\bigskip

The paper is organized as follows.

\bigskip

In Sec. \ref{p-n-tomogram}, we give a brief review of the
photon-number tomogram and its generalization to the two-mode
case. Here we also suggest a method of obtaining the joint
probability distribution corresponding to a two-qubit tomogram. In
Sec. \ref{Schr-cat-example}, an example of Schr\"{o}dinger cat
state is studied in detail. In Sec. \ref{Gaussian-state-tomogram},
we introduce the general approach to deal with the Gaussian
states. A particular example of its application is given and
another way of constructing the qubit portrait is demonstrated. In
Sec. \ref{conclusions}, conclusions and prospects are presented.

\section{\label{p-n-tomogram}Photon-number tomogram and its qubit portrait}

The conventional photon-number tomogram for a one-mode light state
given by the density matrix $\hat{\rho}$ reads
\cite{banaszek,wvogel,mancini}

\begin{equation}
\label{pnt-1} w(n,\alpha) = Tr\left( \hat{\rho}
\hat{D}^{\dag}(\alpha)| n \rangle \langle n | {\hat{D}}(\alpha)
\right) = \langle n | {\hat{D}}(\alpha) \hat{\rho}
\hat{D}^{\dag}(\alpha)| n \rangle,
\end{equation}

\noindent where $|n\rangle$ is an eigenstate of the operator
$\hat{a}^{\dag}\hat{a}$, with $\hat{a}^{\dag}$ and $\hat{a}$ being
the photon creation and annihilation operators, respectively, and
$\hat{D}(\alpha)$ is a displacement operator, which depends on the
complex number $\alpha$ as follows:

\begin{equation}
\label{displacement} \hat{D}(\alpha) = {\rm e}^{\alpha
\hat{a}^{\dag} - \alpha^{\ast} \hat{a}}.
\end{equation}

Thus, the photon-number tomogram is nothing else but the photon
distribution function of the state ${\hat{D}}^{\dag}(\alpha)
\hat{\rho} \hat{D}(\alpha)$. In other words, it is the probability
to find $n$ photons in the state with the amplitude shifted by a
complex number $\alpha$. It is worth noting that once tomogram
(\ref{pnt-1}) is known, the state $\hat{\rho}$ can be
reconstructed. (For Gaussian states the photon distributions were
expressed in terms of Hermite polynomials of several variables in
\cite{dodonov1,dodonov2}).

As far as the two-mode light state is concerned, the formula
(\ref{pnt-1}) changes slightly

\begin{equation}
\label{pnt-2mode} w(n_{1},n_{2},\alpha_{1},\alpha_{2}) = \langle
n_{1}n_{2} | {\hat{D}}(\alpha_{1},\alpha_{2}) \hat{\rho}
\hat{D}^{\dag}(\alpha_{1},\alpha_{2})| n_{1}n_{2} \rangle,
\end{equation}

\noindent where $| n_{1}n_{2} \rangle = | n_{1} \rangle | n_{2}
\rangle$ is the state with $n_{1}$ photons in the first mode and
$n_{2}$ photons in the second mode, and
$\hat{D}(\alpha_{1},\alpha_{2}) = \hat{D}_{1}(\alpha_{1})
\hat{D}_{2}(\alpha_{2})$. Here $\hat{D}_{i}(\alpha _{i})$ denotes
the displacement operator (\ref{displacement}), where we replaced
$\alpha \rightarrow \alpha_{i}$, $\hat{a}^{\dag} \rightarrow
\hat{a}_{i}^{\dag}$, and $\hat{a} \rightarrow \hat{a}_{i}$, with
$\hat{a}_{i}^{\dag}$ and $\hat{a}_{i}$ being the photon creation
and annihilation operators of the $i$th mode, $i=1,2$. The
operators $\hat{D}_{1}(\alpha_{1})$ and $ \hat{D}_{2}(\alpha_{2})$
obviously commute.

The photon-number tomogram (\ref{pnt-1}) is rather similar to the
spin tomogram \cite{dodonovPLA,oman'ko-jetp} of a qudit (particle
with spin $j$). The number of photons $n$ plays the role of the
spin projection $m$ ($m = -j, -j+1,...,j$), while the complex
number $\alpha$ is analogues to the unitary matrix of the
corresponding rotation group. In contrast to the spin tomogram,
the photon-number tomogram has infinite outputs, so there are
infinite ways to construct a qubit portrait \cite{chernega,lupo1}
of such a tomogram. Indeed, if one has the infinite
probability-distribution vector with nonnegative components

\begin{equation}
\overrightarrow{W}_{\infty} = \left(%
\begin{array}{c}
  w(0,\alpha) \\
  w(1,\alpha) \\
  w(2,\alpha) \\
  \ldots \\
\end{array}%
\right),
\end{equation}

\noindent where $\sum\limits_{n=1}^{\infty}{w(n,\alpha)} = 1$, a
new probability-distribution vector $\overrightarrow{W}_{2}$ with
two components can be constructed as follows

\begin{equation}
\overrightarrow{W}_{2} = \left(%
\begin{array}{c}
  \sum\limits_{n\in A}{w(n,\alpha)} \\
  \sum\limits_{n\in (Z_{+}\backslash A)}{w(n,\alpha)} \\
\end{array}%
\right),
\end{equation}

\noindent where $A$ is an arbitrary subset of the set of
nonnegative integers $Z_{+} \equiv \{0,1,2,...\}$. (We denote the
set of nonnegative integers that do not belong to $A$ as
$Z_{+}\backslash A$; from this it follows that $A \cup
(Z_{+}\backslash A) = Z_{+}$, $A \cap (Z_{+}\backslash A) =
\emptyset$).

Let us now consider the photon-number tomogram of the two-mode
light.

Once the parameters $\alpha_{1}$ and $\alpha_{2}$ are given the
function $w(n_{1},n_{2},\alpha_{1},\alpha_{2})$ can be treated as
a table with an infinite number of both rows and columns

\begin{equation}
\label{matrix-probability}
\overline{W}_{\infty,\infty} = \left(%
\begin{array}{cccc}
  w(0,0,\alpha_{1},\alpha_{2}) & w(0,1,\alpha_{1},\alpha_{2}) & w(0,2,\alpha_{1},\alpha_{2}) & \ldots \\
  w(1,0,\alpha_{1},\alpha_{2}) & w(1,1,\alpha_{1},\alpha_{2}) & w(1,2,\alpha_{1},\alpha_{2}) & \ldots \\
  w(2,0,\alpha_{1},\alpha_{2}) & w(2,1,\alpha_{1},\alpha_{2}) & w(2,2,\alpha_{1},\alpha_{2}) & \ldots \\
  \ldots & \ldots & \ldots & \ldots \\
\end{array}%
\right),
\end{equation}

\begin{figure}
\begin{center}
\includegraphics{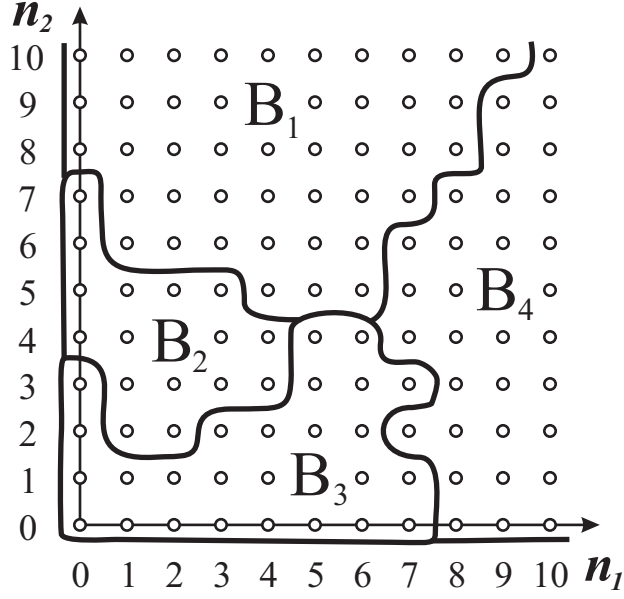}
\caption{\label{plot1}Partition of the set $Z_{+}\times Z_{+}$ to
subsets $B_{i}$, $i=1,...,4$, that enables one to construct the
four-vector (\ref{W-4-wrong}). Nevertheless, such a vector is not
a probability distribution vector of two qubits because $B_{i}$
cannot be written in the form $A_{1}^{(i)}\times A_{2}^{(i)}$,
where $A_{1}^{(i)}$ and $A_{2}^{(i)}$ are subsets of $Z_{+}$.}
\end{center}
\end{figure}

\noindent with the sum
$\sum\limits_{n_{1},n_{2}=0}^{\infty}{w(n_{1},n_{2},\alpha_{1},\alpha_{2})}$
being equal to unity. This distribution induces the
four-dimensional probability distribution vector of the form

\begin{equation}
\label{W-4-wrong}
\overrightarrow{W'}_{4} = \left(%
\begin{array}{c}
  \sum\limits_{(n_{1},n_{2})\in B_{1}}{w(n_{1},n_{2},\alpha_{1},\alpha_{2})} \\
  \sum\limits_{(n_{1},n_{2})\in B_{2}}{w(n_{1},n_{2},\alpha_{1},\alpha_{2})} \\
  \sum\limits_{(n_{1},n_{2})\in B_{3}}{w(n_{1},n_{2},\alpha_{1},\alpha_{2})} \\
  \sum\limits_{(n_{1},n_{2})\in B_{4}}{w(n_{1},n_{2},\alpha_{1},\alpha_{2})} \\
\end{array}%
\right),
\end{equation}

\noindent where the sets $B_{i}$ ($i=1,...,4$) satisfy the
conditions: $B_{i}\subset Z_{+}\times Z_{+}$,
$\bigcup\limits_{i=1}^{4}{B_{i}} = Z_{+}\times Z_{+}$, and
$B_{i}\cap B_{j} = \emptyset$ if $i\neq j$. An example of
separation of the set $Z_{+}\times Z_{+}$ that meets all the above
requirements is illustrated in Fig. \ref{plot1}.

The probability vector obtained can be interpreted readily as the
spin tomogram of a particle with spin $j=3/2$, but it is hardly
able to describe the two-qubit system. The matter is that the
simply separable state of two qubits has the tomogram of a
factorized form
$w(m_{1},m_{2},\overrightarrow{N}_{1},\overrightarrow{N}_{2}) =
w(m_{1},\overrightarrow{N}_{1})w(m_{2},\overrightarrow{N}_{2})$,
where $m_{1}$ and $m_{2}$ are the spin projections of the first
and second spin-$1/2$ particles to the axes
$\overrightarrow{N}_{1}$ and $\overrightarrow{N}_{2}$,
respectively. This feature underlies the Bell-CHSH inequality and
is of a great importance. For the factorization above to be valid,
it requires $B_{i}$ to be the direct product of subsets of
$Z_{+}$, $i=1,...,4$. Consequently, once purposing to construct
the qubit portrait, one should rewrite formula (\ref{W-4-wrong})
in the form

\begin{equation}
\label{W-4}
\overrightarrow{W}_{4} = \left(%
\begin{array}{c}
  \sum\limits_{(n_{1},n_{2})\ \in \ A_{1}\times A_{2}}{w(n_{1},n_{2},\alpha_{1},\alpha_{2})} \\
  \sum\limits_{(n_{1},n_{2})\ \in \ A_{1}\times (Z_{+}\backslash A_{2})}{w(n_{1},n_{2},\alpha_{1},\alpha_{2})} \\
  \sum\limits_{(n_{1},n_{2})\ \in \ (Z_{+}\backslash A_{1})\times A_{2}}{w(n_{1},n_{2},\alpha_{1},\alpha_{2})} \\
  \sum\limits_{(n_{1},n_{2})\ \in \ (Z_{+}\backslash A_{1})\times (Z_{+}\backslash A_{2})}{w(n_{1},n_{2},\alpha_{1},\alpha_{2})} \\
\end{array}%
\right),
\end{equation}

\noindent where $A_{1}$ and $A_{2}$ are subsets of the set of
nonnegative integers $Z_{+}$. If the state is simply separable,
i.e.,

\begin{equation}
w(n_{1},n_{2},\alpha_{1},\alpha_{2}) = w_{1} (n_{1},\alpha_{1})
w_{2} (n_{2},\alpha_{2}),
\end{equation}

\noindent then vector (\ref{W-4}) can be written in factorized
form

\begin{equation}
\label{W-4-factorization}
\left(%
\begin{array}{c}
  \sum\limits_{n_{1}\in A_{1}}{w(n_{1},\alpha_{1}}) \\
  \sum\limits_{n_{1}\in (Z_{+}\backslash A_{1})}{w(n_{1},\alpha_{1}}) \\
\end{array}%
\right) \otimes \left(%
\begin{array}{c}
  \sum\limits_{n_{2}\in A_{2}}{w(n_{2},\alpha_{2}}) \\
  \sum\limits_{n_{2}\in (Z_{+}\backslash A_{2})}{w(n_{2},\alpha_{2}}) \\
\end{array}%
\right),
\end{equation}

\noindent which corresponds to the probability-distribution vector
of the simply separable two-qubit state. As far as separable
states are concerned, the four-vector (\ref{W-4}) becomes merely
the convex sum of the vectors (\ref{W-4-factorization}).

To draw an analogy with the probability-distribution vector of two
qubits, we designate the components of the vector (\ref{W-4}) as
follows:

\begin{equation}
\label{W-4-qubitlike}
\overrightarrow{W}_{4} = \left(%
\begin{array}{c}
  w(+,+,\alpha_{1},\alpha_{2}) \\
  w(+,-,\alpha_{1},\alpha_{2}) \\
  w(-,+,\alpha_{1},\alpha_{2}) \\
  w(-,-,\alpha_{1},\alpha_{2}) \\
\end{array}%
\right),
\end{equation}

\noindent with all the components being nonnegative and their sum
equal to unity.

{\textbf{Comment}}. $\nabla$ The construction of the
four-dimensional probability distribution vector from the matrix
(\ref{matrix-probability}) can be presented in an elegant matrix
form\footnote{Such a map for finite probability vectors was used
in the work of M. A. Man'ko, V. I. Man'ko, and R. V. Mendes
(unpublished)}. To start, we show how one can obtain a
probability-distribution vector $\overrightarrow{w}_{2}$ with two
components with the help of a probability-distribution vector

\begin{equation}
\overrightarrow{w}_{3} = \left(%
\begin{array}{c}
  a \\
  b \\
  c \\
\end{array}%
\right).
\end{equation}

\noindent The general form of transformation
$\overrightarrow{w}_{3}\rightarrow \overrightarrow{w}_{2}$ reads

\begin{equation}
\overrightarrow{w}_{2} = \left(%
\begin{array}{c}
  K \\
  L \\
\end{array}%
\right) = \left(%
\begin{array}{c}
  p_{1}a+p_{2}b+p_{3}c \\
  1-(p_{1}a+p_{2}b+p_{3}c) \\
\end{array}%
\right) = \left(%
\begin{array}{ccc}
  p_{1} & p_{2} & p_{3} \\
  1-p_{1} & 1-p_{2} & 1-p_{3} \\
\end{array}%
\right) \left(%
\begin{array}{c}
  a \\
  b \\
  c \\
\end{array}%
\right),
\end{equation}

\noindent where $p_{i} \ge 0$ and $\sum{p_{i}}=1$. Regarding
$\overrightarrow{w}_{2}$ as a three-dimensional probability
vector, one can write

\begin{equation}
\left(%
\begin{array}{c}
  \overrightarrow{w}_{2} \\
  0 \\
\end{array}%
\right) = \left(%
\begin{array}{c}
  K \\
  L \\
  0 \\
\end{array}%
\right) = \left(%
\begin{array}{ccc}
  p_{1} & p_{2} & p_{3} \\
  1-p_{1} & 1-p_{2} & 1-p_{3} \\
  0 & 0 & 0 \\
\end{array}%
\right)\left(%
\begin{array}{c}
  a \\
  b \\
  c \\
\end{array}%
\right) = M \overrightarrow{w}_{3},
\end{equation}

\noindent where $M$ is a stochastic matrix with no more than two
nonzero rows. A generalization of this method to higher dimensions
is obvious, and one has

\begin{equation}
\left(%
\begin{array}{c}
  \overrightarrow{w}_{2} \\
  0 \\
  ... \\
  0 \\
\end{array}%
\right) = M_{\infty} \overrightarrow{w}_{\infty}.
\end{equation}

\noindent To deal with the probability-distribution matrices, one
can use their vector representation \cite{chernega} or,
equivalently, apply the technic above separately to the rows and
columns. In the case of matrix (\ref{matrix-probability}), we have

\begin{equation}
\left(%
\begin{array}{ccccc}
  w_{++} & w_{+-} & 0 & ... & 0 \\
  w_{-+} & w_{--} & 0 & ... & 0 \\
  0 & 0 & 0 & ... & 0 \\
  ... & ... & ... & ... & ... \\
  0 & 0 & 0 & ... & 0 \\
\end{array}%
\right) = M_{\infty}^{(1)} \overline{W}_{\infty,\infty}
{M_{\infty}^{(2)}}^{T},
\end{equation}

\noindent where the block $\left(%
\begin{array}{cc}
  w_{++} & w_{+-} \\
  w_{-+} & w_{--} \\
\end{array}%
\right)$ provides a required qubit portrait. This completes the
Comment. $\triangle$

Since the four-vector (\ref{W-4-qubitlike}) is merely a
vector-function of parameters $\alpha_{1}$ and $\alpha_{2}$, we
can introduce new complex variables $\beta_{1}$ and $\beta_{2}$
and construct the stochastic matrix of the form

\begin{equation}
\label{M}
M\left(\alpha_{1},\alpha_{2},\beta_{1},\beta_{2}\right) = \left(%
\begin{array}{cccc}
  w(+,+,\alpha_{1},\alpha_{2}) & w(+,+,\alpha_{1},\beta_{2}) & w(+,+,\beta_{1},\alpha_{2}) & w(+,+,\beta_{1},\beta_{2}) \\
  w(+,-,\alpha_{1},\alpha_{2}) & w(+,-,\alpha_{1},\beta_{2}) & w(+,-,\beta_{1},\alpha_{2}) & w(+,-,\beta_{1},\beta_{2}) \\
  w(-,+,\alpha_{1},\alpha_{2}) & w(-,+,\alpha_{1},\beta_{2}) & w(-,+,\beta_{1},\alpha_{2}) & w(-,+,\beta_{1},\beta_{2}) \\
  w(-,-,\alpha_{1},\alpha_{2}) & w(-,-,\alpha_{1},\beta_{2}) & w(-,-,\beta_{1},\alpha_{2}) & w(-,-,\beta_{1},\beta_{2}) \\
\end{array}%
\right).
\end{equation}

The matrix (\ref{M}) is obtained just in the same way as its
analog for a two-qubit system \cite{andreev-shchukin}. For this
reason it exhibits the same properties. The most essential point
is the following. If the initial state of two-mode light is
separable then, the Bell-CHSH inequality is fulfilled for certain.
That implies that the inequality

\begin{equation}
\label{Bell-CHSH} \biggl| \textrm{Tr} \Bigl(
M(\alpha_{1},\alpha_{2},\beta_{1},\beta_{2}) \ I \Bigr) \biggr|
\leq 2
\end{equation}

\noindent holds true, where the matrix $I$ is given by the formula

\begin{equation}
\label{I-matrix}
I = \left(%
\begin{array}{cccc}
  1 & -1 & -1 & 1 \\
  1 & -1 & -1 & 1 \\
  1 & -1 & -1 & 1 \\
  -1 & 1 & 1 & -1 \\
\end{array}%
\right).
\end{equation}

If there exist complex numbers $\alpha_{1}$, $\alpha_{2}$,
$\beta_{1}$, $\beta_{2}$ such that the inequality
(\ref{Bell-CHSH}) is violated, the two-mode light state is
entangled. Consequently, fulfilling the Bell-CHSH inequality is a
necessary condition of separability of the two-mode light state.

\section{\label{Schr-cat-example}Detecting entanglement of the Schr\"{o}dinger cat state}

To demonstrate the application of the technique described above,
we consider the state that is entangled by construction. (One
cannot help mentioning that the entanglement of the
Schr\"{o}dinger cat states was detected by means of symplectic
tomograms in \cite{kolesnikov-mipt}). Here we investigate the
Schr\"{o}dinger cat state of the form
\cite{NadeemAnsari-Manko,dodonov-nikonov}

\begin{equation}
\label{cat-state} |\psi\rangle = N(\gamma_{1},\gamma_{2})\Bigl(|
\gamma_{1}, \gamma_{2} \rangle + | -\gamma_{1}, -\gamma_{2}
\rangle\Bigr),
\end{equation}

\noindent where $| \gamma_{1}, \gamma_{2} \rangle = | \gamma_{1}
\rangle | \gamma_{2} \rangle$, states $| \gamma_{i} \rangle$ are
eigenstates of the photon annihilation operators, i.e.,
$\hat{a}_{i} | \gamma_{i} \rangle = \gamma_{i} | \gamma_{i}
\rangle$, $i=1,2$ (coherent states of the first and the second
modes, respectively). The factor $N(\gamma_{1},\gamma_{2})$ can be
found by employing the normalization requirement $\langle \psi |
\psi \rangle = 1$. Indeed, since for arbitrary coherent states
$|\gamma\rangle$ and $|\delta\rangle$ the scalar product $\langle
\gamma | \delta \rangle$ can be written in the form
\cite{sudarshan-63,glauber}

\begin{equation}
\label{Schr-state} \langle \gamma | \delta \rangle =
\exp\left\{-\frac{1}{2}{|\gamma|}^{2}
-\frac{1}{2}{|\delta|}^{2}+\gamma^{\ast}\delta\right\},
\end{equation}

\noindent it is of no difficulty to calculate

\begin{eqnarray}
\langle \psi | \psi \rangle = N^{2}(\gamma_{1},\gamma_{2})\Bigl(
\langle \gamma_{1}, \gamma_{2} | \gamma_{1}, \gamma_{2} \rangle +
2{\rm{Re}}(\langle \gamma_{1}, \gamma_{2} | -\gamma_{1},
-\gamma_{2} \rangle) + \langle -\gamma_{1}, -\gamma_{2} |
-\gamma_{1}, -\gamma_{2} \rangle
\Bigr)\nonumber\\
 = 2 N^{2}(\gamma_{1},\gamma_{2}) \left( 1 +
{\rm e}^{ - 2{|\gamma_{1}|}^{2} - 2{|\gamma_{2}|}^{2}}\right) = 4
N^{2}(\gamma_{1},\gamma_{2}) {\rm e}^{ - \left({|\gamma_{1}|}^{2}
+ {|\gamma_{2}|}^{2}\right)} \cosh\left( {|\gamma_{1}|}^{2} +
{|\gamma_{2}|}^{2} \right).
\end{eqnarray}

\noindent Therefore,

\begin{equation}
\label{Normalization} N(\gamma_{1},\gamma_{2}) = \frac{\exp\left\{
\frac{1}{2} \left({|\gamma_{1}|}^{2} +
{|\gamma_{2}|}^{2}\right)\right\}}{2\left\{\cosh\left(
{|\gamma_{1}|}^{2} + {|\gamma_{2}|}^{2} \right)\right\}^{1/2}}.
\end{equation}

In case of pure quantum state, i.e. $\hat{\rho} = |\psi\rangle
\langle\psi|$, the formula (\ref{pnt-2mode}) takes a more
convenient form

\begin{equation}
\label{pnt-pure} w_{\psi}(n_{1},n_{2},\alpha_{1},\alpha_{2}) =
{\left| \langle n_{1}n_{2} | {\hat{D}}(\alpha_{1},\alpha_{2}) |
\psi \rangle \right|}^{2}.
\end{equation}

To get the explicit expression of the photon-number tomogram in
our particular case, we recall that $|\gamma\rangle =
\hat{D}(\gamma)|0\rangle$ and $\hat{D}(\alpha)\hat{D}(\gamma) =
\hat{D}(\alpha + \gamma){\rm
e}^{(\alpha\gamma^{\ast}-\alpha^{\ast}\gamma)/2}$. This implies
that the displacement operator transforms any coherent state into
a coherent one. To be more precise,

\begin{equation}
\label{D-alpha-plus-gamma}
\hat{D}(\alpha_{1},\alpha_{2})|\gamma_{1}, \gamma_{2}\rangle =
\hat{D}_{1}(\alpha_{1})|\gamma_{1}\rangle
\hat{D}_{2}(\alpha_{2})|\gamma_{2}\rangle = {\rm
e}^{\left(\alpha_{1}\gamma^{\ast}_{1}-\alpha^{\ast}_{1}\gamma_{1}
+\alpha_{2}\gamma^{\ast}_{2}-\alpha^{\ast}_{2}\gamma_{2}\right)/2}|\alpha_{1}
+\gamma_{1}\rangle|\alpha_{2}+\gamma_{2}\rangle.
\end{equation}

\noindent In much the same way, one can write

\begin{equation}
\label{D-alpha-minus-gamma}
\hat{D}(\alpha_{1},\alpha_{2})|-\gamma_{1}, -\gamma_{2}\rangle =
\hat{D}_{1}(\alpha_{1})|-\gamma_{1}\rangle
\hat{D}_{2}(\alpha_{2})|-\gamma_{2}\rangle = {\rm
e}^{\left(\alpha^{\ast}_{1}\gamma_{1}-\alpha_{1}\gamma^{\ast}_{1}
+\alpha^{\ast}_{2}\gamma_{2}-\alpha_{2}\gamma^{\ast}_{2}\right)/2}
|\alpha_{1}-\gamma_{1}\rangle|\alpha_{2}-\gamma_{2}\rangle.
\end{equation}

\noindent The scalar product of the coherent state
$|\delta\rangle$ and the Fock basis state $|n\rangle$ is

\begin{equation}
\label{n-delta} \langle n|\delta\rangle = \langle n|\left({\rm
e}^{-|\delta|^{2}/2}\sum\limits_{k=0}^{\infty}
{\frac{\delta^{k}}{\sqrt{k!}}|k\rangle}\right) = {\rm
e}^{-|\delta|^{2}/2} \frac{\delta^{n}}{\sqrt{n!}}.
\end{equation}

\noindent Using the simplified formula (\ref{pnt-pure}) and taking
into account the results of Eqs. (\ref{Schr-state}),
(\ref{Normalization}), (\ref{D-alpha-plus-gamma}),
(\ref{D-alpha-minus-gamma}), and (\ref{n-delta}), we arrive at the
following formula for the photon-number tomogram of the state
under consideration

\begin{eqnarray}
\label{cat-explicit}
\lefteqn{w_{SC}(n_{1},n_{2},\alpha_{1},\alpha_{2}) = \frac{{\rm
e}^{-\left(|\alpha_{1}|^{2}+|\alpha_{2}|^{2}\right)}}{4
n_{1}!n_{2}!
\cosh\left(|\gamma_{1}|^{2}+|\gamma_{2}|^{2}\right)}} \nonumber \\
& &{}\times\left| {\rm
e}^{-\left(\alpha_{1}^{\ast}\gamma_{1}+\alpha_{2}^{\ast}\gamma_{2}\right)}
(\alpha_{1}+\gamma_{1})^{n_{1}}(\alpha_{2}+\gamma_{2})^{n_{2}} +
{\rm e}^{\alpha_{1}^{\ast}\gamma_{1}+\alpha_{2}^{\ast}\gamma_{2}}
(\alpha_{1}-\gamma_{1})^{n_{1}}(\alpha_{2}-\gamma_{2})^{n_{2}}
\right|^{2}.
\end{eqnarray}

\begin{figure}
\begin{center}
\includegraphics{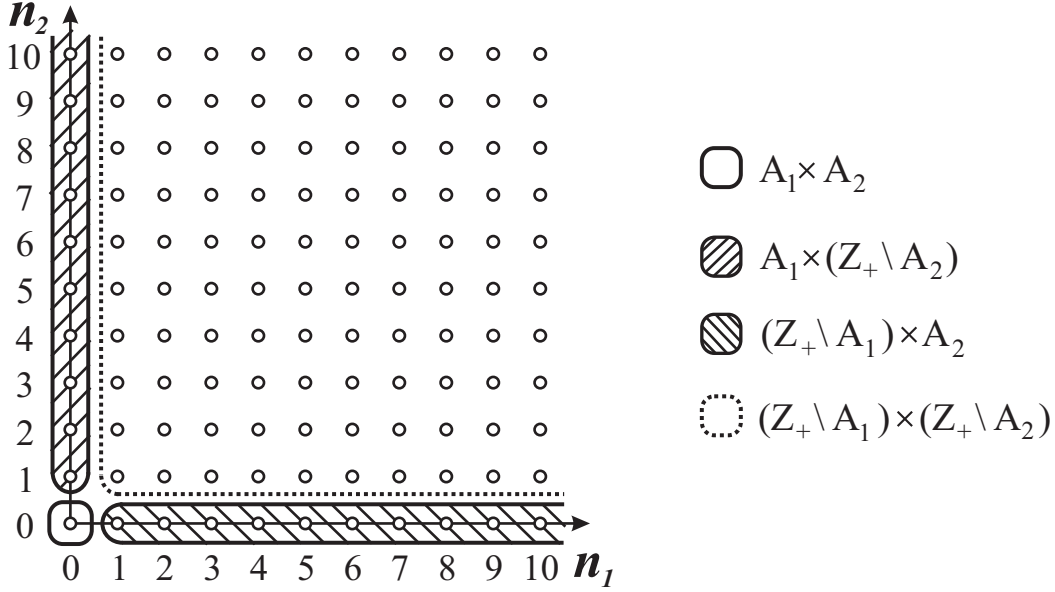}
\caption{\label{zero-nonzero} Simplest example of dividing the set
$Z_{+}\times Z_{+}$ into parts, that enables one to construct the
qubit portrait of the photon-number tomogram.}
\end{center}
\end{figure}

Now, when the photon-number tomogram is known, our goal is to make
sure that the Bell-CHSH inequality can be violated, because the
state involved is entangled. In Sec. \ref{p-n-tomogram}, we
emphasized that there exist many ways to construct the portrait
corresponding to the two-qubit system. In our opinion, the
simplest way is to split the tomogram components corresponding to
the states:

(i) the vacuum state,

(ii) states with no photons in the first mode and with nonzero
number of photons in the second one,

(iii) states with nonzero number of photons in the first mode and
with no photons in the second one,

(iv) states with nonzero number of photons in both modes.

\noindent Then the probability distribution vector (\ref{W-4})
takes the form

\begin{equation}
\label{W-4-cat}
\overrightarrow{W}_{4\ SC} (\alpha_{1},\alpha_{2})= \left(%
\begin{array}{c}
  w(0,0,\alpha_{1},\alpha_{2}) \\
  \sum\limits_{n_{2}=1}^{\infty}{w(0,n_{2},\alpha_{1},\alpha_{2})} \\
  \sum\limits_{n_{1}=1}^{\infty}{w(n_{1},0,\alpha_{1},\alpha_{2})} \\
  \sum\limits_{n_{1},n_{2}=1}^{\infty}{w(n_{1},n_{2},\alpha_{1},\alpha_{2})} \\
\end{array}%
\right) = \left(%
\begin{array}{c}
  w_{SC}(+,+,\alpha_{1},\alpha_{2}) \\
  w_{SC}(+,-,\alpha_{1},\alpha_{2}) \\
  w_{SC}(-,+,\alpha_{1},\alpha_{2}) \\
  w_{SC}(-,-,\alpha_{1},\alpha_{2}) \\
\end{array}%
\right).
\end{equation}

\noindent The pattern of such a partition is shown in Fig.
\ref{zero-nonzero}. Using the explicit expression
(\ref{cat-explicit}) of the photon-number tomogram for our
particular case, we get the components of of the four-vector
(\ref{W-4-cat})

\begin{equation}
w_{SC}(+,+,\alpha_{1},\alpha_{2}) =
\frac{e^{-\left(|\alpha_{1}|^{2}+|\alpha_{2}|^{2}\right)}}{2
\cosh\left(|\gamma_{1}|^{2}+|\gamma_{2}|^{2}\right)} \biggl\{
\cosh
\Bigl(2\textrm{Re}(\alpha_{1}\gamma_{1}+\alpha_{2}\gamma_{2})
\Bigr)+ \cos
\Bigl(2\textrm{Im}(\alpha_{1}\gamma_{1}+\alpha_{2}\gamma_{2})
\Bigr) \biggr\},
\end{equation}

\begin{eqnarray}
\lefteqn{w_{SC}(+,-,\alpha_{1},\alpha_{2}) =
\frac{e^{-\left(|\alpha_{1}|^{2}+|\alpha_{2}|^{2}\right)}}{2
\cosh\left(|\gamma_{1}|^{2}+|\gamma_{2}|^{2}\right)} \biggl\{
e^{|\alpha_{2}|^{2}+|\gamma_{2}|^{2}}\cosh \Bigl(
2\textrm{Re}(\alpha_{1}\gamma_{1}) \Bigr)} \nonumber \\
& &{}+ e^{|\alpha_{2}|^{2}-|\gamma_{2}|^{2}}\cos \Bigl(
2\textrm{Im}(\alpha_{1}\gamma_{1}) \Bigr) - \cosh \Bigl(
2\textrm{Re}(\alpha_{1}\gamma_{1}+\alpha_{2}\gamma_{2}) \Bigr)-
\cos \Bigl(
2\textrm{Im}(\alpha_{1}\gamma_{1}+\alpha_{2}\gamma_{2}) \Bigr)
\biggr\},
\end{eqnarray}

\begin{eqnarray}
\lefteqn{w_{SC}(-,+,\alpha_{1},\alpha_{2}) =
\frac{e^{-\left(|\alpha_{1}|^{2}+|\alpha_{2}|^{2}\right)}}{2
\cosh\left(|\gamma_{1}|^{2}+|\gamma_{2}|^{2}\right)} \biggl\{
e^{|\alpha_{1}|^{2}+|\gamma_{1}|^{2}}\cosh\Bigl(2\textrm{Re}(\alpha_{2}\gamma_{2})\Bigr)} \nonumber \\
& &{}
+e^{|\alpha_{1}|^{2}-|\gamma_{1}|^{2}}\cos\Bigl(2\textrm{Im}(\alpha_{2}\gamma_{2})\Bigr)
-\cosh\Bigl(2\textrm{Re}(\alpha_{1}\gamma_{1}+\alpha_{2}\gamma_{2})\Bigr)-
\cos\Bigl(2\textrm{Im}(\alpha_{1}\gamma_{1}+\alpha_{2}\gamma_{2})\Bigr)
\biggr\},
\end{eqnarray}

\begin{equation}
w_{SC}(-,-,\alpha_{1},\alpha_{2}) = 1
-w_{SC}(+,+,\alpha_{1},\alpha_{2})
-w_{SC}(+,-,\alpha_{1},\alpha_{2})
-w_{SC}(-,+,\alpha_{1},\alpha_{2}).
\end{equation}

Let us now consider the function of four complex variables (or
eight real variables)

\begin{equation}
\label{Bell-like-number}
B_{SC}(\alpha_{1},\alpha_{2},\beta_{1},\beta_{2}) =
\biggl|\textrm{Tr}\Bigl(M_{SC}(\alpha_{1},\alpha_{2},\beta_{1},\beta_{2})\
I\Bigr)\biggr|,
\end{equation}

\noindent where the matrix
$M_{SC}(\alpha_{1},\alpha_{2},\beta_{1},\beta_{2})$ is given by
the formula

\begin{equation}
\label{Matrix-M-Bell-CHSH}
M_{SC}(\alpha_{1},\alpha_{2},\beta_{1},\beta_{2}) = \left(%
\begin{array}{cccc}
  \overrightarrow{W}_{4\ SC} (\alpha_{1},\alpha_{2})
  & \overrightarrow{W}_{4\ SC} (\alpha_{1},\beta_{2})
  & \overrightarrow{W}_{4\ SC} (\beta_{1},\alpha_{2})
  & \overrightarrow{W}_{4\ SC} (\beta_{1},\beta_{2}) \\
\end{array}%
\right),
\end{equation}

\begin{figure}
\begin{center}
\includegraphics{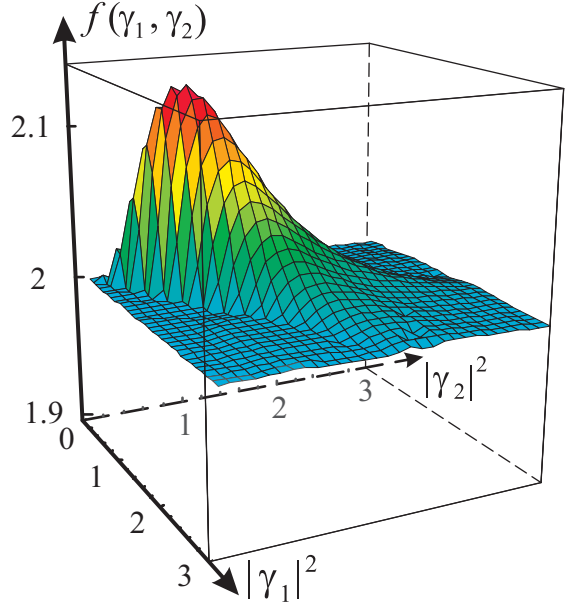}
\caption{\label{MaxFunction} Violation of the Bell-CHSH inequality
for the Schr\"{o}dinger cat states of the form
$|\gamma_{1},\gamma_{2}\rangle+|-\gamma_{1},-\gamma_{2}\rangle$.
Here the maximum value of the Bell number is denoted as
$f(\gamma_{1},\gamma_{2})$. The qubit portrait is constructed
using formula (\ref{W-4-cat}) ("zero-nonzero" approach).}
\end{center}
\end{figure}

\noindent and the matrix $I$ is determined by Eq.
(\ref{I-matrix}).

To demonstrate that there exist such values of variables
$\alpha_{1}$, $\alpha_{2}$, $\beta_{1}$, and $\beta_{2}$ such that
the value of function
$B_{SC}(\alpha_{1},\alpha_{2},\beta_{1},\beta_{2})$ is greater
than 2, we find the maximum of this function, which depends
entirely on the form of the quantum state

\begin{equation}
f(\gamma_{1},\gamma_{2}) =
\max\limits_{\alpha_{1},\alpha_{2},\beta_{1},\beta_{2} \in
\rm{C}}{B_{SC}(\alpha_{1},\alpha_{2},\beta_{1},\beta_{2})}.
\end{equation}

Numerical calculations show that, whatever values of complex
variables $\gamma_{1}$ and $\gamma_{2}$ are taken, the value of
function $f(\gamma_{1},\gamma_{2})$ is never less than 2. This
fact is depicted in Fig. \ref{MaxFunction}. Nevertheless, the
extent to which $f(\gamma_{1},\gamma_{2})$ is greater than 2
differs from one state to the other. If $|\gamma_{1}| \gg 1$ and
$|\gamma_{2}| \gg 1$, then the efficiency of the technique
proposed essentially decreases.

\section{\label{Gaussian-state-tomogram}Detecting entanglement of the Gaussian states}

The next example to be investigated is a nonclassical state of
light described by the Gaussian density matrix (see, e.g.,
\cite{simon,chirkin-3,illuminati,werner}). It is of some interest
because it can characterize a two-mode mixed squeezed state. A
Gaussian density operator $\hat{\rho}$ of the two-mode light is
described by the Wigner function $W(p_{1},p_{2},q_{1},q_{2})$ of
the form \cite{dodonov1}

\begin{equation}
\label{Wigner-2-mode} W(p_{1},p_{2},q_{1},q_{2}) =
\frac{1}{\sqrt{\det\bf{M}}}\exp\left\{-\frac{1}{2}(\bf{Q}-\langle
\bf{Q} \rangle)\bf{M}^{-1}(\bf{Q}-\langle \bf{Q} \rangle)\right\},
\end{equation}

\noindent where

\begin{equation}
\bf{Q}=\left(%
\begin{array}{c}
  p_{1} \\
  p_{2} \\
  q_{1} \\
  q_{2} \\
\end{array}%
\right),\ \ \ \ \ \ \ \ \
\bf{M}=\left(%
\begin{array}{cccc}
  \sigma_{p_{1}p_{1}} & \sigma_{p_{1}p_{2}} & \sigma_{p_{1}q_{1}} & \sigma_{p_{1}q_{2}} \\
  \sigma_{p_{2}p_{1}} & \sigma_{p_{2}p_{2}} & \sigma_{p_{2}q_{1}} & \sigma_{p_{2}q_{2}} \\
  \sigma_{q_{1}p_{1}} & \sigma_{q_{1}p_{2}} & \sigma_{q_{1}q_{1}} & \sigma_{q_{1}q_{2}} \\
  \sigma_{q_{2}p_{1}} & \sigma_{q_{2}p_{2}} & \sigma_{q_{2}q_{1}} & \sigma_{q_{2}q_{2}} \\
\end{array}%
\right).
\end{equation}

\noindent Here averaging $\langle A\rangle$ implies $\langle
A\rangle = {\rm{Tr}} \hat{\rho} \hat{A}$, where $\hat{A}$ can be
one of the operators $\hat{p}_{j} =
-i(\hat{a}_{j}-\hat{a}_{j}^{\dag})/\sqrt{2}$, $\hat{q}_{j} =
(\hat{a}_{j}+\hat{a}_{j}^{\dag})/\sqrt{2}$, $j=1,2$ (variables are
assumed to be dimensionless, with $\hbar = 1$). The components of
the real symmetric dispersion matrix $\bf{M}$ read

\begin{equation}
\label{M-elements} \sigma_{p_{i}p_{j}} = \langle
\hat{p}_{i}\hat{p}_{j}\rangle - \langle \hat{p}_{i}\rangle\langle
\hat{p}_{j}\rangle,\ \ \sigma_{q_{i}q_{j}} = \langle
\hat{q}_{i}\hat{q}_{j}\rangle - \langle \hat{q}_{i}\rangle\langle
\hat{q}_{j}\rangle,\ \ \sigma_{p_{i}q_{j}} = \sigma_{q_{j}p_{i}} =
\frac{1}{2}\langle
\hat{p}_{i}\hat{q}_{j}+\hat{q}_{j}\hat{p}_{i}\rangle - \langle
\hat{p}_{i}\rangle\langle \hat{q}_{j}\rangle.
\end{equation}

Thus, all the information on the state is contained in the Wigner
function (\ref{Wigner-2-mode}) depending on 14 real variables
(four of them determine the vector $\langle \bf{Q} \rangle$ and
ten of them define the matrix $\bf{M}$).

To obtain the photon-number tomogram (\ref{pnt-2mode}), it is
worth noting that this is merely a photon distribution function of
the state
$\hat{D}(\alpha_{1},\alpha_{2})\hat{\rho}\hat{D}^{\dag}(\alpha_{1},\alpha_{2})$.
Let us now reveal what is the Wigner function of such a state. For
this purpose, we calculate the new average values (see, e.g.,
\cite{omanko}). For instance,

\begin{eqnarray}
\langle p_{1}\rangle_{\alpha_{1}\alpha_{2}} &&=
{\rm{Tr}}\left(\hat{D}(\alpha_{1},\alpha_{2})\hat{\rho}\hat{D}^{\dag}(\alpha_{1},\alpha_{2})\hat{p}_{1}\right)
\nonumber \\&&=
{\rm{Tr}}\left(\hat{\rho}\hat{D}_{1}^{\dag}(\alpha_{1})\hat{D}_{2}^{\dag}(\alpha_{2})\left\{-i(\hat{a}_{1}-\hat{a}_{1}^{\dag})/\sqrt{2}\right\}\hat{D}_{1}(\alpha_{1})\hat{D}_{2}(\alpha_{2})\right)
\nonumber\\&&=
{\rm{Tr}}\left(\hat{\rho}\left\{-i(\hat{a}_{1}+\alpha_{1}-\hat{a}_{1}^{\dag}-\alpha_{1}^{\ast})/\sqrt{2}\right\}\right) \nonumber\\
&&={\rm{Tr}}\left(\hat{\rho}\left\{\hat{p}_{1}+\sqrt{2}\
\textrm{Im}\ \alpha_{1}\right\}\right) = \langle
p_{1}\rangle+\sqrt{2}\ \textrm{Im}\ \alpha_{1}.
\end{eqnarray}

\noindent Here we used the relation
$\hat{D}^{\dag}(\alpha)\hat{a}\hat{D}(\alpha) = \hat{a} + \alpha
\hat{1}$. In a similar way we find $\langle p_{2} \rangle
_{\alpha_{1}\alpha_{2}} = \langle p_{2} \rangle + \sqrt{2}\
\textrm{Im}\ \alpha_{2}$, $\langle q_{1} \rangle
_{\alpha_{1}\alpha_{2}} = \langle q_{1} \rangle + \sqrt{2} \
\textrm{Re}\ \alpha_{1}$, and $\langle q_{2} \rangle
_{\alpha_{1}\alpha_{2}} = \langle q_{2} \rangle + \sqrt{2}\
\textrm{Re}\ \alpha_{2}$. The remarkable fact is that the elements
of the matrix $\bf{M}$ do not change through a transition from
$\hat{\rho}$ to
$\hat{D}(\alpha_{1},\alpha_{2})\hat{\rho}\hat{D}^{\dag}(\alpha_{1},\alpha_{2})$.
It follows directly from the formulas (\ref{M-elements}). We will
refer to the vector of new average values as
$\langle{\bf{Q}}\rangle_{\alpha_{1}\alpha_{2}}$.

The photon distribution function of a generic $N$-mode mixed
Gaussian state of light has been obtained earlier
\cite{omanko,dodonov1,dodonov2}. We use it to compute the
photon-number tomogram (\ref{pnt-2mode}). The result is

\begin{equation}
\label{pnt-Gauss} w_{G}(n_{1},n_{2},\alpha_{1},\alpha_{2}) =
\frac{\exp\left\{-\langle {\bf{Q}} \rangle_{\alpha_{1}\alpha_{2}}
(2{\bf{M}}+{\bf{I}}_{4})^{-1} \langle {\bf{Q}}
\rangle_{\alpha_{1}\alpha_{2}}\right\}}{\sqrt{\det\left({\bf{M}} +
\frac{1}{2} {\bf{I}}_{4}\right)}} \ \frac
{H_{n_{1},n_{2},n_{1},n_{2}}^{\{\bf{R}\}}({\bf{y})}} { n_{1}!\
n_{2}!},
\end{equation}

\noindent where ${{\bf{I}}_{4}}$ is the $4\times 4$ identity
matrix, and the four-dimensional matrix $\bf{R}$ and the
four-vector $\bf{y}$ can be expressed with the help of the matrix

\begin{equation}
{\bf{U}} = \frac{1}{\sqrt{2}} \left(%
\begin{array}{cccc}
  -i & 0 & i & 0 \\
  0 & -i & 0 & i \\
  1 & 0 & 1 & 0 \\
  0 & 1 & 0 & 1 \\
\end{array}%
\right)
\end{equation}

\noindent as

\begin{eqnarray}
\label{R-matrix} {\bf{R}} = {\bf{U}}^{\dag}
\left({\bf{I}}_{4}-2{\bf{M}}\right)
\left({\bf{I}}_{4}+2{\bf{M}}\right)^{-1} {\bf{U}}^{\ast},\\
\label{y-vector} {\bf{y}} = 2{\bf{U}}^{tr}
\left({\bf{I}}_{4}-2{\bf{M}}\right)^{-1}
\langle{\bf{Q}}\rangle_{\alpha_{1}\alpha_{2}},
\end{eqnarray}

\noindent and the term of the form
$H_{k_{1},k_{2},k_{3},k_{4}}^{\{\bf{R}\}}(\bf{x})$ is the
four-dimensional Hermite polynomial given by the formula
\cite{fian}

\begin{equation}
H_{k_{1}, k_{2}, k_{3}, k_{4}}^{\{\bf{R}\}}({\bf{x}}) =
(-1)^{k_{1}+k_{2}+k_{3}+k_{4}} \
\exp\left(\frac{1}{2}{\bf{xRx}}\right) \
\frac{\partial^{k_{1}+k_{2}+k_{3}+k_{4}}} {\partial x_{1}^{k_{1}}
\partial x_{2}^{k_{2}} \partial x_{3}^{k_{3}} \partial
x_{4}^{k_{4}}} \ \exp\left( - \frac{1}{2} \sum \limits_{i,j=1}^{4}
{R_{ij}x_{i}x_{j}}\right).
\end{equation}

Once one has the expression of the photon-number tomogram
(\ref{pnt-Gauss}), it seems possible to explore the entanglement
of the Gaussian states just in the same way as in Sec.
\ref{Schr-cat-example}. By analogy with the previous example, we
choose the state which is entangled for sure. For instance, the
squeezed state given by its Wigner function of the form

\begin{equation}
\label{Gauss-ent-state-example} W(p_{1},p_{2},q_{1},q_{2}) = 4
\exp\left\{-2 \left( 3p_{1}^{2} - \sqrt{35}p_{1}p_{2} + 3p_{2}^{2}
+ q_{1}^{2} -\sqrt{3}q_{1}q_{2} + q_{2}^{2}\right)\right\}
\end{equation}

\noindent is entangled, because it contains the products
$p_{1}p_{2}$ and $q_{1}q_{2}$ in the exponent. This state
corresponds to the matrix $\bf{M}$ and the vector
$\langle{\bf{Q}}\rangle$ of the form

\begin{equation}
{\bf{M}} = \left(%
\begin{array}{cccc}
  3 & \sqrt{35}/2 & 0 & 0 \\
  \sqrt{35}/2 & 3 & 0 & 0 \\
  0 & 0 & 1 & \sqrt{3}/2 \\
  0 & 0 & \sqrt{3}/2 & 1 \\
\end{array}%
\right), \ \ \ \ \ \langle{\bf{Q}}\rangle = \left(%
\begin{array}{c}
  0 \\
  0 \\
  0 \\
  0 \\
\end{array}%
\right),
\end{equation}

\noindent which meets the generalized uncertainty relation $\det
{\bf{M}} \ge 1/4^{2}$. The substitution of $\bf{M}$ for such a
matrix in Eqs. (\ref{R-matrix}) and (\ref{y-vector}) yields

\begin{equation}
{\bf{R}} = \frac{1}{42}\left(%
\begin{array}{cccc}
  0 & 3\sqrt{35}-7\sqrt{3} & 0 & -3\sqrt{35}-7\sqrt{3} \\
  3\sqrt{35}-7\sqrt{3} & 0 & -3\sqrt{35}-7\sqrt{3} & 0 \\
  0 & -3\sqrt{35}-7\sqrt{3} & 0 & 3\sqrt{35}-7\sqrt{3} \\
  -3\sqrt{35}-7\sqrt{3} & 0 & 3\sqrt{35}-7\sqrt{3}& 0 \\
\end{array}%
\right)\;
\end{equation}

\noindent and

\begin{equation}
{\bf{y}} = \left(%
\begin{array}{c}
  -i(\alpha_{1}-\sqrt{\frac{7}{5}}\ \textrm{Re}\ \alpha_{2})-\sqrt{3}\ \textrm{Im}\ \alpha_{2}\\
  -i(\alpha_{2}-\sqrt{\frac{7}{5}}\ \textrm{Re}\ \alpha_{1})-\sqrt{3}\ \textrm{Im}\ \alpha_{1}\\
  i(\alpha_{1}^{\ast}-\sqrt{\frac{7}{5}}\ \textrm{Re}\ \alpha_{2})-\sqrt{3}\ \textrm{Im}\ \alpha_{2}\\
  i(\alpha_{2}^{\ast}-\sqrt{\frac{7}{5}}\ \textrm{Re}\ \alpha_{1})-\sqrt{3}\ \textrm{Im}\ \alpha_{1}\\
\end{array}%
\right).
\end{equation}

\bigskip

Now the photon-number tomogram (\ref{pnt-Gauss}) can be calculated
for any given parameters $n_{1}$, $n_{2}$, $\alpha_{1}$, and
$\alpha_{2}$. It is just the right time for trying to detect the
entanglement of the state (\ref{Gauss-ent-state-example}). To
start, we apply the approach developed above, which means
constructing the probability-distribution vector (\ref{W-4-cat}),
and then the stochastic matrix (\ref{Matrix-M-Bell-CHSH}), the
trace of the product of which with the matrix (\ref{I-matrix})
leads to the Bell-like number. The obstacle arises during the
first step, because the probability-distribution four-vector
(\ref{W-4-cat}) is hardly computable analytically. The numerical
calculation has some particular features concerned with
four-dimensional Hermite polynomials. The matter is that it is
rather difficult to calculate
$H_{n_{1},n_{2},n_{1},n_{2}}^{\{\bf{R}\}}({\bf{y})}$ when
$n_{1}\gg 1$ and $n_{2}\gg 1$. Therefore, we deal with numbers
$n_{1}$ and $n_{2}$ confined by the ranges $0 \le n_{1} \le 30$,
$0 \le n_{2} \le 30$. To neglect the values of
$w_{G}(n_{1},n_{2},\alpha_{1},\alpha_{2})$ providing $n_{1} > 30$
or $n_{2} > 30$ we impose the limitations $|\textrm{Re}\alpha_{i}|
\le 2$, $|\textrm{Im}\alpha_{i}| \le 2$, $i=1,2$ (and consequently
$|\textrm{Re}\beta_{i}| \le 2$, $|\textrm{Im}\beta_{i}| \le 2$,
$i=1,2$). If this is the case, the total contribution of
probabilities $w(n_{1},n_{2},\alpha_{1},\alpha_{2})$ with high
photon numbers ($n_{1}>30$, $n_{2}>30$) to the distribution vector
(\ref{W-4-cat}) does not exceed $10^{-4}$ and serves as an error
measure. The calculation shows that the Bell-like number
(\ref{Bell-like-number}) is less than 2 providing the variables
$\alpha_{1}$, $\alpha_{2}$, $\beta_{1}$, and $\beta_{2}$ are not
beyond the region specified above.

\bigskip

The problem occurs not due to restrictions introduced but because
of the unlucky choice of the qubit-portrait construction. Indeed,
during the maximization process the variables $\alpha_{1}$,
$\alpha_{2}$, $\beta_{1}$, and $\beta_{2}$ attempt to distribute
in such a way that the quantity
$w_{G}(n_{1},n_{2},\alpha_{1},\alpha_{2})$ is far from zero when
$n_{1} \approx n_{2}\approx 0$, the quantity
$w_{G}(n_{1},n_{2},\alpha_{1},\beta_{2})$ is far from zero when
$n_{1} \approx 0$ and $n_{2}\gg 1$, the quantity
$w_{G}(n_{1},n_{2},\beta_{1},\alpha_{2})$ is far from zero when
$n_{1} \gg 1$ and $n_{2} \approx 0$, and the quantity
$w_{G}(n_{1},n_{2},\beta_{1},\beta_{2})$ is far from zero when
$n_{1} \gg 1$ and $n_{2} \gg 1$. In Fig. \ref{zero-nonzero}, one
can see that the narrowness of the domains $B_{2}$ and $B_{3}$
prevents the quantities
$\sum\limits_{n_{2}=1}^{\infty}{w(0,n_{2},\alpha_{1},\beta_{2})}$
and
$\sum\limits_{n_{1}=1}^{\infty}{w(n_{1},0,\beta_{1},\alpha_{2})}$
from increasing. In our case they cannot be greater then $2/3$.
(See analogues problem discussed in \cite{dowling1})

\begin{figure}
\begin{center}
\includegraphics{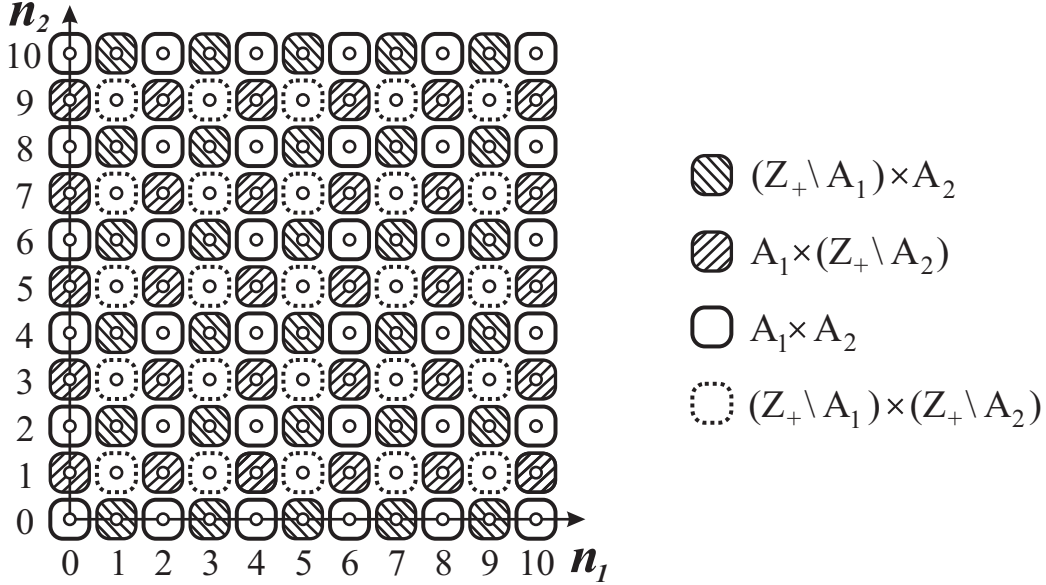}
\caption{\label{New-partition} Alternative partition of the set
$Z_{+}\times Z_{+}$. The sets $A_{1}$ and $A_{2}$ are nonnegative
even integers, while the sets $Z_{+}\backslash A_{1}$ and
$Z_{+}\backslash A_{2}$ are nonnegative odd integers.}
\end{center}
\end{figure}

\bigskip

To put the situation right we suggest the other separation
$\{A_{1}\times A_{2}, A_{1}\times (Z_{+}\backslash A_{2}),
(Z_{+}\backslash A_{1})\times A_{2}, (Z_{+}\backslash A_{1})\times
(Z_{+}\backslash A_{2})\}$ of the set $Z_{+}\times Z_{+}$, which
is needed to construct the four-dimensional
probability-distribution vector. The main idea of the partitioning
proposed is patterned in Fig. \ref{New-partition}. It implies
distinguishing even and odd numbers of photons in individual
modes. Such a choice is stimulated by the peculiarities of the
photon-distribution function of squeezed states concerned with its
oscillations \cite{kolesnikov,schrade}. Thus, the explicit
expression of the four-vector $\overrightarrow{M}_{4\ G}$ reads

\begin{equation}
\label{W-4-Gauss}
\overrightarrow{W}_{4\ G} (\alpha_{1},\alpha_{2})= \left(%
\begin{array}{c}
  \sum\limits_{n_{1}=0}^{\infty}{\sum\limits_{n_{2}=0}^{\infty}{w(2n_{1},2n_{2},\alpha_{1},\alpha_{2})}} \\
  \sum\limits_{n_{1}=0}^{\infty}{\sum\limits_{n_{2}=0}^{\infty}{w(2n_{1},2n_{2}+1,\alpha_{1},\alpha_{2})}} \\
  \sum\limits_{n_{1}=0}^{\infty}{\sum\limits_{n_{2}=0}^{\infty}{w(2n_{1}+1,2n_{2},\alpha_{1},\alpha_{2})}} \\
  \sum\limits_{n_{1}=0}^{\infty}{\sum\limits_{n_{2}=0}^{\infty}{w(2n_{1}+1,2n_{2}+1,\alpha_{1},\alpha_{2})}} \\
\end{array}%
\right) = \left(%
\begin{array}{c}
  w_{G}(+,+,\alpha_{1},\alpha_{2}) \\
  w_{G}(+,-,\alpha_{1},\alpha_{2}) \\
  w_{G}(-,+,\alpha_{1},\alpha_{2}) \\
  w_{G}(-,-,\alpha_{1},\alpha_{2}) \\
\end{array}%
\right).
\end{equation}

\noindent The matrix
$M_{G}(\alpha_{1},\alpha_{2},\beta_{1},\beta_{2})$ appears
naturally in the same way as the matrix
(\ref{Matrix-M-Bell-CHSH}). To demonstrate that the inequality
$|Tr(M_{G}(\alpha_{1},\alpha_{2},\beta_{1},\beta_{2})I)| \le 2$
can be violated, it is enough to make the substitution
$\alpha_{1}=-0.12i$, $\alpha_{2}=0.04i$, $\beta_{1}=0.22i$,
$\beta_{2}=-0.32i$. Then the matrix $M_{G}$ takes the form

\begin{equation}
M_{G}(-0.12i,0.04i,0.22i,-0.32i) = \left(%
\begin{array}{cccc}
  0.6199 & 0.5907 & 0.6083 & 0.4678 \\
  0.0222 & 0.0515 & 0.0291 & 0.1696 \\
  0.0241 & 0.0395 & 0.0357 & 0.1624 \\
  0.3335 & 0.3181 & 0.3266 & 0.2000 \\
\end{array}%
\right),
\end{equation}

\noindent and $|Tr(M_{G}(-0.12i,0.04i,0.22i,-0.32i)I)| \approx
2.26 > 2$. The violation of the Bell-CHSH inequality for this
particular set of variables $\alpha_{1}$, $\alpha_{2}$,
$\beta_{1}$, and $\beta_{2}$ indicates immediately that the
Gaussian state (\ref{Gauss-ent-state-example}) is entangled.

\bigskip

While dealing with Gaussian states, one should notice that the
presented method of tomogram calculation works well for both pure
and mixed states. This is accompanied by the complexity of
calculations including multidimensional Hermite polynomial
contrast to the relatively easy ones for pure states. To
demonstrate the advantages of the technique proposed, we consider
the family of generally mixed states given by matrix ${\bf{M}}$
and the vector ${\bf{Q}}$ of the form

\begin{equation}
\label{M-and-Q-k-l}
{\bf{M}}(k,l) = \left(%
\begin{array}{cccc}
  k+\frac{l}{k} & \sqrt{k^{2}-\frac{1}{4}} & 0 & 0 \\
  \sqrt{k^{2}-\frac{1}{4}} & k & 0 & 0 \\
  0 & 0 & k & \sqrt{k^{2}-\frac{1}{4}} \\
  0 & 0 & \sqrt{k^{2}-\frac{1}{4}} & k \\
\end{array}%
\right), \ \ \ \ \ \langle{\bf{Q}}\rangle = \left(%
\begin{array}{c}
  0 \\
  0 \\
  0 \\
  0 \\
\end{array}%
\right).
\end{equation}

\begin{figure}
\begin{center}
\includegraphics{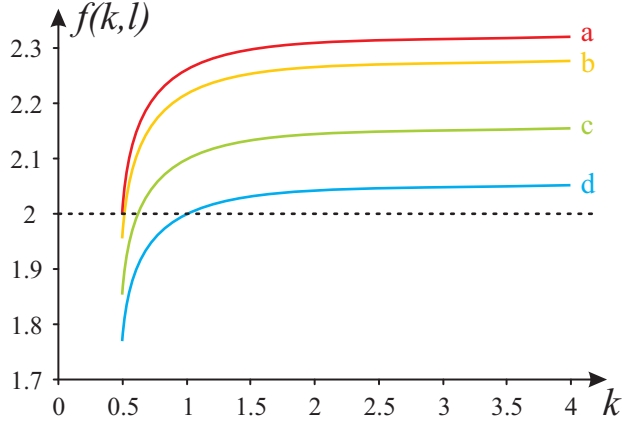}
\caption{\label{plotabcd} Violation of the Bell-CHSH inequality
for the family of states (\ref{M-and-Q-k-l}). The maximum of the
Bell-like number is denoted by $f(k,l)$. The purity of states does
not change while moving along the separate curve: pure states with
$l=0$ (a), mixed states with $l=0.01$ (b), $l=0.04$ (c), and
$l=0.07$ (d).}
\end{center}
\end{figure}

The purity of this state is determined by the parameter $l$ only,
because $\det{\bf{M}}=(1+4l)/16$. Thus the state is pure when
$l=0$ and mixed for all $l>0$. Denoting

\begin{equation}
f(k,l)=\max\limits_{\alpha_{1},\alpha_{2},\beta_{1},\beta_{2}\in{\rm{C}}}{{\rm{Tr}}(B(\alpha_{1},\alpha_{2},\beta_{1},\beta_{2})I)}
\end{equation}

\noindent we succeeded in plotting the dependence of $f(k,l)$ on
$k$ for the states with different purities (see Fig.
\ref{plotabcd}). Though the violation of the Bell-CHSH inequality
is clearly observed for pure states, it is rather difficult to
detect entanglement of mixed states. This fact is similar to the
two-qubit system behaviour when the increase in mixing causes the
loss of entanglement.

\bigskip

We have been just convinced that the technique of constructing the
qubit portrait proposed in this section can be efficient enough
for detecting entanglement of Gaussian states. To check if it
works properly for other kinds of states, we apply it to the
Schr\"{o}dinger cat state discussed in the previous section.
Though the photon-number tomogram of such a state can be obtained
by means of addition of probability distribution functions of
Gaussian states, we are going to exploit the explicit formula
(\ref{cat-explicit}).

\bigskip

Then the even-odd photon number form of the probability
distribution vector (\ref{W-4-Gauss}) reads

\begin{eqnarray}
&w(+,+,\alpha_{1},\alpha_{2}) =  \frac {{\rm
e}^{-\left(|\alpha_{1}|^{2}+|\alpha_{2}|^{2}\right)}}
{4\cosh\left(|\gamma_{1}|^{2}+|\gamma_{2}|^{2}\right)} \times
\Biggl\{ {\rm
e}^{-2{\rm{Re}}\left(\alpha_{1}^{\ast}\gamma_{1}+\alpha_{2}^{\ast}\gamma_{2}\right)}
\cosh\left(|\alpha_{1}+\gamma_{1}|^{2}\right)
\cosh\left(|\alpha_{2}+\gamma_{2}|^{2}\right)\nonumber \\
& \ + {\rm
e}^{2{\rm{Re}}\left(\alpha_{1}^{\ast}\gamma_{1}+\alpha_{2}^{\ast}\gamma_{2}\right)}
\cosh\left(|\alpha_{1}-\gamma_{1}|^{2}\right)
\cosh\left(|\alpha_{2}-\gamma_{2}|^{2}\right)\nonumber \\
& \ +2 \cos\Bigl(2\textrm{Im}(\alpha_{1}^{\ast}\gamma_{1}+\alpha_{2}^{\ast}\gamma_{2})\Bigr)\nonumber \\
& \ \ \ \times \biggl[
\cosh\left(|\alpha_{1}|^{2}-|\gamma_{1}|^{2}\right)
\cosh\left(|\alpha_{2}|^{2}-|\gamma_{2}|^{2}\right)
\cos\Bigl(2{\rm{Im}}(\alpha_{1}^{\ast}\gamma_{1})\Bigr)
\cos\Bigl(2{\rm{Im}}(\alpha_{2}^{\ast}\gamma_{2})\Bigr)\nonumber
\\&  \ \ \ \ \ \ - \sinh\left(|\alpha_{1}|^{2}-|\gamma_{1}|^{2}\right)
\sinh\left(|\alpha_{2}|^{2}-|\gamma_{2}|^{2}\right)
\sin\Bigl(2{\rm{Im}}(\alpha_{1}^{\ast}\gamma_{1})\Bigr)
\sin\Bigl(2{\rm{Im}}(\alpha_{2}^{\ast}\gamma_{2})\Bigr) \biggr]\nonumber \\
& \ +2
\sin\Bigl(2{\rm{Im}}(\alpha_{1}^{\ast}\gamma_{1}+\alpha_{2}^{\ast}\gamma_{2})\Bigr)\nonumber \\
& \ \ \ \times \biggl[
\cosh\left(|\alpha_{1}|^{2}-|\gamma_{1}|^{2}\right)
\sinh\left(|\alpha_{2}|^{2}-|\gamma_{2}|^{2}\right)
\cos\Bigl(2{\rm{Im}}(\alpha_{1}^{\ast}\gamma_{1})\Bigr)
\sin\Bigl(2{\rm{Im}}(\alpha_{2}^{\ast}\gamma_{2})\Bigr)\nonumber
\\& \ \ \ \ \ \ + \sinh\left(|\alpha_{1}|^{2}-|\gamma_{1}|^{2}\right)
\cosh\left(|\alpha_{2}|^{2}-|\gamma_{2}|^{2}\right)
\sin\Bigl(2{\rm{Im}}(\alpha_{1}^{\ast}\gamma_{1})\Bigr)
\cos\Bigl(2{\rm{Im}}(\alpha_{2}^{\ast}\gamma_{2})\Bigr)\biggr]
\Biggr\},
\\&w(+,-,\alpha_{1},\alpha_{2}) =  \frac {{\rm
e}^{-\left(|\alpha_{1}|^{2}+|\alpha_{2}|^{2}\right)}}
{4\cosh\left(|\gamma_{1}|^{2}+|\gamma_{2}|^{2}\right)} \times
\Biggl\{ {\rm
e}^{-2{\rm{Re}}\left(\alpha_{1}^{\ast}\gamma_{1}+\alpha_{2}^{\ast}\gamma_{2}\right)}
\cosh\left(|\alpha_{1}+\gamma_{1}|^{2}\right)
\sinh\left(|\alpha_{2}+\gamma_{2}|^{2}\right)\nonumber \\
& \ + {\rm
e}^{2{\rm{Re}}\left(\alpha_{1}^{\ast}\gamma_{1}+\alpha_{2}^{\ast}\gamma_{2}\right)}
\cosh\left(|\alpha_{1}-\gamma_{1}|^{2}\right)
\sinh\left(|\alpha_{2}-\gamma_{2}|^{2}\right)\nonumber \\
& \ +2 \cos\Bigl(2\textrm{Im}(\alpha_{1}^{\ast}\gamma_{1}+\alpha_{2}^{\ast}\gamma_{2})\Bigr)\nonumber \\
& \ \ \ \times \biggl[
\cosh\left(|\alpha_{1}|^{2}-|\gamma_{1}|^{2}\right)
\sinh\left(|\alpha_{2}|^{2}-|\gamma_{2}|^{2}\right)
\cos\Bigl(2{\rm{Im}}(\alpha_{1}^{\ast}\gamma_{1})\Bigr)
\cos\Bigl(2{\rm{Im}}(\alpha_{2}^{\ast}\gamma_{2})\Bigr)\nonumber
\\&  \ \ \ \ \ \ - \sinh\left(|\alpha_{1}|^{2}-|\gamma_{1}|^{2}\right)
\cosh\left(|\alpha_{2}|^{2}-|\gamma_{2}|^{2}\right)
\sin\Bigl(2{\rm{Im}}(\alpha_{1}^{\ast}\gamma_{1})\Bigr)
\sin\Bigl(2{\rm{Im}}(\alpha_{2}^{\ast}\gamma_{2})\Bigr) \biggr]\nonumber \\
& \ +2
\sin\Bigl(2{\rm{Im}}(\alpha_{1}^{\ast}\gamma_{1}+\alpha_{2}^{\ast}\gamma_{2})\Bigr)\nonumber \\
& \ \ \ \times \biggl[
\cosh\left(|\alpha_{1}|^{2}-|\gamma_{1}|^{2}\right)
\cosh\left(|\alpha_{2}|^{2}-|\gamma_{2}|^{2}\right)
\cos\Bigl(2{\rm{Im}}(\alpha_{1}^{\ast}\gamma_{1})\Bigr)
\sin\Bigl(2{\rm{Im}}(\alpha_{2}^{\ast}\gamma_{2})\Bigr)\nonumber
\\& \ \ \ \ \ \ + \sinh\left(|\alpha_{1}|^{2}-|\gamma_{1}|^{2}\right)
\sinh\left(|\alpha_{2}|^{2}-|\gamma_{2}|^{2}\right)
\sin\Bigl(2{\rm{Im}}(\alpha_{1}^{\ast}\gamma_{1})\Bigr)
\cos\Bigl(2{\rm{Im}}(\alpha_{2}^{\ast}\gamma_{2})\Bigr)\biggr]
\Biggr\},
\\&w(-,+,\alpha_{1},\alpha_{2}) =  \frac {{\rm
e}^{-\left(|\alpha_{1}|^{2}+|\alpha_{2}|^{2}\right)}}
{4\cosh\left(|\gamma_{1}|^{2}+|\gamma_{2}|^{2}\right)} \times
\Biggl\{ {\rm
e}^{-2{\rm{Re}}\left(\alpha_{1}^{\ast}\gamma_{1}+\alpha_{2}^{\ast}\gamma_{2}\right)}
\sinh\left(|\alpha_{1}+\gamma_{1}|^{2}\right)
\cosh\left(|\alpha_{2}+\gamma_{2}|^{2}\right)\nonumber \\
& \ + {\rm
e}^{2{\rm{Re}}\left(\alpha_{1}^{\ast}\gamma_{1}+\alpha_{2}^{\ast}\gamma_{2}\right)}
\sinh\left(|\alpha_{1}-\gamma_{1}|^{2}\right)
\cosh\left(|\alpha_{2}-\gamma_{2}|^{2}\right)\nonumber \\
& \ +2 \cos\Bigl(2\textrm{Im}(\alpha_{1}^{\ast}\gamma_{1}+\alpha_{2}^{\ast}\gamma_{2})\Bigr)\nonumber \\
& \ \ \ \times \biggl[
\sinh\left(|\alpha_{1}|^{2}-|\gamma_{1}|^{2}\right)
\cosh\left(|\alpha_{2}|^{2}-|\gamma_{2}|^{2}\right)
\cos\Bigl(2{\rm{Im}}(\alpha_{1}^{\ast}\gamma_{1})\Bigr)
\cos\Bigl(2{\rm{Im}}(\alpha_{2}^{\ast}\gamma_{2})\Bigr)\nonumber
\\&  \ \ \ \ \ \ - \cosh\left(|\alpha_{1}|^{2}-|\gamma_{1}|^{2}\right)
\sinh\left(|\alpha_{2}|^{2}-|\gamma_{2}|^{2}\right)
\sin\Bigl(2{\rm{Im}}(\alpha_{1}^{\ast}\gamma_{1})\Bigr)
\sin\Bigl(2{\rm{Im}}(\alpha_{2}^{\ast}\gamma_{2})\Bigr) \biggr]\nonumber \\
& \ +2
\sin\Bigl(2{\rm{Im}}(\alpha_{1}^{\ast}\gamma_{1}+\alpha_{2}^{\ast}\gamma_{2})\Bigr)\nonumber \\
& \ \ \ \times \biggl[
\cosh\left(|\alpha_{1}|^{2}-|\gamma_{1}|^{2}\right)
\cosh\left(|\alpha_{2}|^{2}-|\gamma_{2}|^{2}\right)
\sin\Bigl(2{\rm{Im}}(\alpha_{1}^{\ast}\gamma_{1})\Bigr)
\cos\Bigl(2{\rm{Im}}(\alpha_{2}^{\ast}\gamma_{2})\Bigr)\nonumber
\\& \ \ \ \ \ \ + \sinh\left(|\alpha_{1}|^{2}-|\gamma_{1}|^{2}\right)
\sinh\left(|\alpha_{2}|^{2}-|\gamma_{2}|^{2}\right)
\cos\Bigl(2{\rm{Im}}(\alpha_{1}^{\ast}\gamma_{1})\Bigr)
\sin\Bigl(2{\rm{Im}}(\alpha_{2}^{\ast}\gamma_{2})\Bigr)\biggr]
\Biggr\},
\\&w(-,-,\alpha_{1},\alpha_{2}) = \frac {{\rm
e}^{-\left(|\alpha_{1}|^{2}+|\alpha_{2}|^{2}\right)}}
{4\cosh\left(|\gamma_{1}|^{2}+|\gamma_{2}|^{2}\right)} \times
\Biggl\{ {\rm
e}^{-2{\rm{Re}}\left(\alpha_{1}^{\ast}\gamma_{1}+\alpha_{2}^{\ast}\gamma_{2}\right)}
\sinh\left(|\alpha_{1}+\gamma_{1}|^{2}\right)
\sinh\left(|\alpha_{2}+\gamma_{2}|^{2}\right)\nonumber \\
& \ + {\rm
e}^{2{\rm{Re}}\left(\alpha_{1}^{\ast}\gamma_{1}+\alpha_{2}^{\ast}\gamma_{2}\right)}
\sinh\left(|\alpha_{1}-\gamma_{1}|^{2}\right)
\sinh\left(|\alpha_{2}-\gamma_{2}|^{2}\right)\nonumber \\
& \ +2 \cos\Bigl(2\textrm{Im}(\alpha_{1}^{\ast}\gamma_{1}+\alpha_{2}^{\ast}\gamma_{2})\Bigr)\nonumber \\
& \ \ \ \times \biggl[
\sinh\left(|\alpha_{1}|^{2}-|\gamma_{1}|^{2}\right)
\sinh\left(|\alpha_{2}|^{2}-|\gamma_{2}|^{2}\right)
\cos\Bigl(2{\rm{Im}}(\alpha_{1}^{\ast}\gamma_{1})\Bigr)
\cos\Bigl(2{\rm{Im}}(\alpha_{2}^{\ast}\gamma_{2})\Bigr)\nonumber
\\&  \ \ \ \ \ \ - \cosh\left(|\alpha_{1}|^{2}-|\gamma_{1}|^{2}\right)
\cosh\left(|\alpha_{2}|^{2}-|\gamma_{2}|^{2}\right)
\sin\Bigl(2{\rm{Im}}(\alpha_{1}^{\ast}\gamma_{1})\Bigr)
\sin\Bigl(2{\rm{Im}}(\alpha_{2}^{\ast}\gamma_{2})\Bigr) \biggr]\nonumber \\
& \ +2
\sin\Bigl(2{\rm{Im}}(\alpha_{1}^{\ast}\gamma_{1}+\alpha_{2}^{\ast}\gamma_{2})\Bigr)\nonumber \\
& \ \ \ \times \biggl[
\sinh\left(|\alpha_{1}|^{2}-|\gamma_{1}|^{2}\right)
\cosh\left(|\alpha_{2}|^{2}-|\gamma_{2}|^{2}\right)
\cos\Bigl(2{\rm{Im}}(\alpha_{1}^{\ast}\gamma_{1})\Bigr)
\sin\Bigl(2{\rm{Im}}(\alpha_{2}^{\ast}\gamma_{2})\Bigr)\nonumber
\\& \ \ \ \ \ \ + \cosh\left(|\alpha_{1}|^{2}-|\gamma_{1}|^{2}\right)
\sinh\left(|\alpha_{2}|^{2}-|\gamma_{2}|^{2}\right)
\sin\Bigl(2{\rm{Im}}(\alpha_{1}^{\ast}\gamma_{1})\Bigr)
\cos\Bigl(2{\rm{Im}}(\alpha_{2}^{\ast}\gamma_{2})\Bigr)\biggr]
\Biggr\}.
\end{eqnarray}

\begin{figure}
\begin{center}
\includegraphics{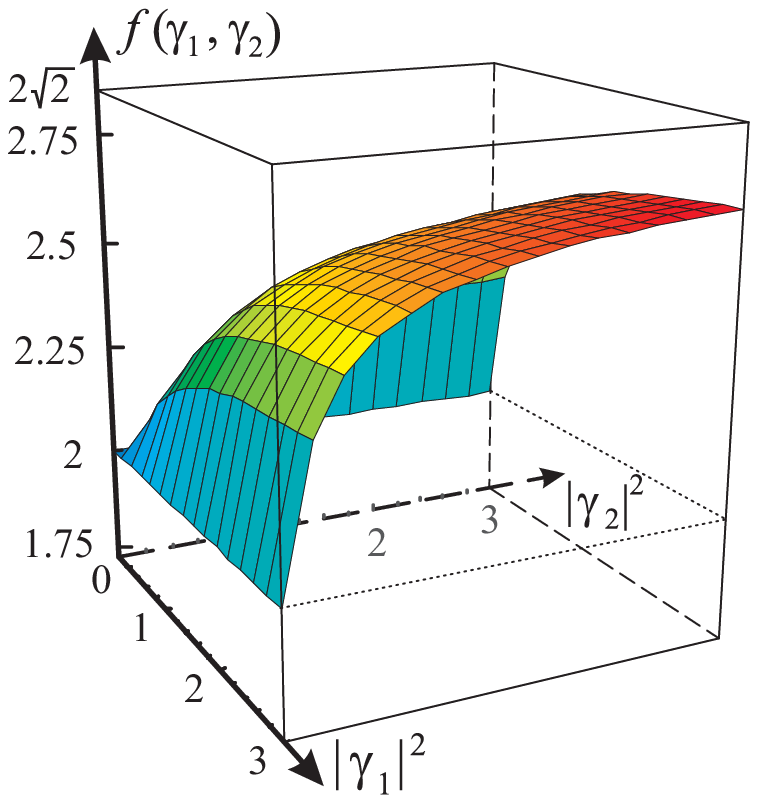}
\caption{\label{MaxFunction-EvenOdd} Violation of the Bell-CHSH
inequality for the Schr\"{o}dinger cat states of the form
$|\gamma_{1},\gamma_{2}\rangle+|-\gamma_{1},-\gamma_{2}\rangle$.
Here the maximum value of the Bell-like number is denoted by
$f(\gamma_{1},\gamma_{2})$. The qubit portrait is constructed
using formula (\ref{W-4-Gauss}) ("even-odd" approach).}
\end{center}
\end{figure}

By a method analogues to that used in Sec. \ref{Schr-cat-example},
we calculate the maximum of the Bell-like number
$B_{SC}(\alpha_{1},\alpha_{2},\beta_{1},\beta_{2}) =
|Tr(M_{SC}(\alpha_{1},\alpha_{2},\beta_{1},\beta_{2})I)|$ over the
complex numbers $\alpha_{1}$, $\alpha_{2}$, $\beta_{1}$, and
$\beta_{2}$, which is a function depending entirely on the form of
the quantum state

\begin{equation}
f(\gamma_{1},\gamma_{2}) =
\max\limits_{\alpha_{1},\alpha_{2},\beta_{1},\beta_{2} \in
C}{B_{SC}(\alpha_{1},\alpha_{2},\beta_{1},\beta_{2})}.
\end{equation}

The numerical calculation of this function for real numbers
$\gamma_{1}$ and $\gamma_{2}$ is shown in Fig.
\ref{MaxFunction-EvenOdd}. Comparison of Fig.
\ref{MaxFunction-EvenOdd} with Fig. \ref{MaxFunction} apparently
shows that the "even-odd" approach of constructing the qubit
portrait achieves better results than the "zero-nonzero" method.

\bigskip

In Fig. \ref{MaxFunction-EvenOdd}, we can see that the
Schr\"{o}dinger cat state (\ref{cat-state}) is always entangled
except the cases $\gamma_{1}=0$ or $\gamma_{2}=0$, where it
becomes separable
$|0\rangle(|\gamma_{2}\rangle+|-\gamma_{2}\rangle)$ or
$(|\gamma_{1}\rangle+|-\gamma_{1}\rangle)|0\rangle$. While moving
to large $|\gamma_{1}|$ and $|\gamma_{1}|$, the function
$f(\gamma_{1},\gamma_{2})$ increases. This increase can be
ascribed to the fact that if $|\gamma_{i}| \gg 1$, the states
$|\gamma_{i}\rangle$ and $|-\gamma_{i}\rangle$ become
quasiorthogonal due to $|\langle\gamma_{i}|-\gamma_{i}\rangle|\ll
1$. If this is the case, the state (\ref{cat-state}) resembles the
usual two-qubit cat state
$\frac{1}{\sqrt{2}}(|00\rangle+|11\rangle)$, which is maximally
entangled and leads to the greatest violation of the Bell
inequality $B=2\sqrt{2}$ (also known as the Cirelson bound
\cite{cirelson}). As far as our case of two-mode light state is
concerned, the function $f(\gamma_{1},\gamma_{2})$ equals
approximately $2.7$ when $\gamma_{1}=\gamma_{2}=10$, equals
approximately $2.78$ when $\gamma_{1}=\gamma_{2}=50$, and seems to
tend to $2\sqrt{2}$ when $|\gamma_{1}|$ and $|\gamma_{2}|$ go to
infinity.

\section{\label{conclusions}Conclusions}

The photon-number tomogram is similar to the qudit tomogram (i.e.,
spin tomogram of the system with the total spin $j$), because of
the countability of the outputs. The conventional approach of
constructing the qubit portrait of qudit states can be applied to
the pnoton number tomogram with some modifications. Being as
powerful as its qudit counterpart, such a method suits for
detecting entanglement of two-mode light states. To be more exact,
the violation of the Bell-CHSH inequality indicates immediately
that the state is entangled. Since there are many ways of reducing
the photon-number tomogram of the two-mode light state to the spin
tomogram of the two-qubit system, the effectiveness of the
entanglement detection depends on the method utilized. To
illustrate this fact we compared two of them in this paper.

\section*{Acknowledgments}
V. I. M. thanks the Russian Foundation for Basic Research for
partial support under Project Nos. 07-02-00598 and 08-02-90300.

\end{document}